\documentclass[iop]{emulateapj}
\usepackage{epsfig}

\usepackage[usenames,dvips]{color}

\shorttitle{Multi-wavelength emissions of PSR J1023+0038}
\shortauthors{Takata et al.}

\newcommand{\flux}{\,erg\,cm$^{-2}$\,s$^{-1}$}

\newcommand{\lum}{\,erg\,s$^{-1}$}

\newcommand{\cm}{\,cm$^{-2}$}
\newcommand{\nh}{$N_\mathrm{H}$}

\begin{document}
\slugcomment{Accepted for publication in ApJ}
\title{Multi-wavelength emissions from the millisecond pulsar binary PSR~J1023+0038 during an accretion active state}

\author{J. Takata\altaffilmark{1}, K.~L. Li\altaffilmark{2},
Gene C.~K. Leung\altaffilmark{1}, A.~K.~H. Kong\altaffilmark{2},
 P.~H.~T. Tam\altaffilmark{2}, C.~Y. Hui\altaffilmark{3},
E.~M.~H. Wu\altaffilmark{1}, Yi Xing\altaffilmark{4,2}, 
 Yi Cao\altaffilmark{5}, Sumin Tang\altaffilmark{5,6}, Zhongxiang Wang\altaffilmark{4}, and K.~S. Cheng\altaffilmark{1}}

\affil{ $^1$ Department of Physics, University of Hong Kong,
Pokfulam Road, Hong Kong; takata@hku.hk} 
 \affil{$^2$ Institute of Astronomy and Department of Physics, National Tsing Hua University, Hsinchu 30013, Taiwan; akong@phys.nthu.edu.tw}
\affil{$^3$ 
Department of Astronomy and Space Science, Chungnam National University, Daejeon, Republic of Korea}  
\affil{$^4$ Shanghai Astronomical Observatory, Chinese Academy of Sciences, 80 Nandan Road, Shanghai 200030, China}
\affil{$^5$ Astronomy Department, California Institute of Technology, 1200 E. California Boulevard, Pasadena, CA 91125, USA}
\affil{$^6$ Kavli Institute for Theoretical Physics, University of California, Santa Barbara, CA 93106, USA}



\begin{abstract}
Recent observations strongly suggest that the millisecond pulsar binary PSR J1023+0038 has developed an accretion disk since 2013 June.
We present a multi-wavelength analysis of PSR J1023+0038, which reveals 
that 1) its gamma-rays suddenly brightened within a few days in June/July 2013 and has remained at a high gamma-ray state for several months; 2) both UV and X-ray fluxes have increased by roughly an order of 
magnitude, and 3) the spectral energy distribution has changed significantly after the gamma-ray sudden flux change. Time variabilities associated with UV and X-rays are on the order of
100-500 seconds and 50-100 seconds, respectively.
 Our model suggests that a newly formed accretion disk due to
 the sudden increase of the stellar wind could
 explain the changes of all these observed features. The increase of UV is 
emitted from the disk, and a new component in gamma-rays is produced by inverse 
Compton scattering between the new UV component and pulsar wind. The 
increase of X-rays results from the enhancement of injection pulsar wind 
energy into the intra-binary shock due to the increase of the stellar wind. 
We also predict that the radio pulses may be blocked by the evaporated winds from the disk and the pulsar is still powered by rotation.
\end{abstract}


\section{Introduction}
The formation and evolution of millisecond pulsars (MSPs) have been a
subject of debate. It is widely believed that neutron stars in
low-mass X-ray binaries (LMXBs) can be spun up to millisecond periods
by gaining angular momentum from the accreting materials of the
companion star (e.g. Alpar et al. 1982). When the accretion stops, the
neutron stars become radio millisecond pulsars powered by rotation
(e.g. Campana et al. 1998). The first direct observational evidence to
support the link between MSPs and LMXBs comes from the discovery of
the X-ray binary/millisecond pulsar PSR J1023+0038 (J1023 hereafter).
J1023 was first identified as an LMXB with an orbital period of 4.75
hours based on its X-ray and optical properties (Woudt et al. 2004;
Thorstensen \& Armstrong 2005; Homer et al. 2006) and the millisecond
pulsation ($P=1.69$ms) was found subsequently (Archibald et al. 2009).
The source clearly showed an accretion disk between 2000 and 2001
(Wang et al. 2009) and the disk has since disappeared (Archibald et
al. 2009) while radio pulsation was found in 2007 confirming the MSP
nature (Archibald et al. 2009). Therefore, J1023 is considered as a
newly born MSP, representing the missing link of a rotation-powered
MSP descended from an LMXB. In such a system that recently transited
from the LMXB phase to the radio MSP phase, it is widely believed that
the accretion disk of J1023 will reform due to an increase in mass
transfer rate, and the radio MSP will then be switched off (Archibald
et al. 2009; Wang et al. 2009; Tam et al. 2010). One very recent
example is IGR J18245--2452 in the globular cluster M28 for which it
has been switching between an accretion and a rotation-powered MSP
(Papitto et al. 2013).

Since 2013 late-June, the radio pulsation of J1023 has disappeared
between 350 MHz and 5 GHz (Stappers et al. 2013; Patruno et al. 2014).
Meanwhile, recent optical spectroscopy shows strong double peaked
H$\alpha$ emission indicating that an accretion disk is formed
(Halpern et al. 2013). Moreover, the X-ray emission has increased by a
factor of at least 20 comparing with previous quiescent values (Kong
2013) and the UV emission has brightened by 4 magnitudes (Patruno et
al. 2014). All these strongly indicate that J1023 is switching from a
radio MSP to a LMXB with an accretion disk. In addition, the gamma-ray
emission ($>100$ MeV) as seen with {\it Fermi} has been reported to brighten by a factor of 5 (Stappers et al. 2013; Patruno et al. 2014). It is
somewhat unusual as we expect that the gamma-rays should be turned off
like radio during the accretion state (e.g., Takata et al. 2010,2012).
In fact, no LMXB has been detected by {\it Fermi} so far.

In this paper, we report a multi-wavelength (from optical to
gamma-ray) study of J1023 during its recent accretion active 
state (\S~\ref{observation}),
and propose a model to explain the multi-wavelength 
behaviors (\S~\ref{theory}). A brief summary and discussion on 
observed  variabilities in multi-wavelength emissions  
are given in \S~4.

\section{Summary of multi-wavelength observations}
\label{observation}
\subsection{{\it Fermi Large Area Telescope} $\gamma$-ray observations}
Gamma-ray data were obtained, reduced and analyzed using the {\it Fermi} 
Science Tools package (v9r32p5), available from the {\it Fermi} Science 
Support Center \footnote{http://fermi.gsfc.nasa.gov/ssc/data/analysis/software/}. We selected events in the reprocessed Pass 7 'Source' class and used the
 P7REP\_SOURCE\_V15 version of the instrumental response functions. To reduce
 contamination from the Earth's albedo, we excluded time intervals when the
 ROI was observed at zenith angles greater than 100\degr\ or when the rocking
 angle of the LAT was greater than 52\degr. We used photons between 0.1 and 
300 GeV within a $20\degr\times20\degr$ ROI centered at the position of 
J1023. The Fermi Large Area Telescope (LAT) was under the sky survey mode for most of the time considered, allowing the full sky to be surveyed every three hours.

We performed binned likelihood analyzes with the \emph{gtlike} tool. For source 
modeling, all 2FGL catalog sources (Nolan et al., 2012) within $19\degr$ of 
the ROI center, the galactic diffuse emission (gll\_iem\_v05.fits), and 
isotropic diffuse emission (iso\_source\_v05.txt) were included. For sources
 more than 10\degr\ away from the position of J1023, the spectral 
parameters were fixed to the catalog values.

\subsubsection{Spectral Analysis}
We used {\it Fermi} LAT data collected in two separate 
time intervals: from August 14, 2008 to May 31, 2013 and from July 1, 2013 to 
November 12, 2013, roughly corresponding to the time before and after the $\gamma$-ray state change (see below for a temporal analysis),
 respectively. 
In each of the two time intervals, we performed two separate fits. We modeled 
J1023 with a simple power law
\begin{equation}
\frac{dN}{dE} = N_0 \left(\frac{E}{E_0}\right)^{-\Gamma},
\end{equation}
as well as a power law with exponential cutoff (PLE)
\begin{equation}
\frac{dN}{dE} = N_0 \left(\frac{E}{E_0}\right)^{-\Gamma}\exp(-\frac{E}{E_c}).
\end{equation}

Before the state change, the fit with a simple power law gives $\Gamma = 2.6 \pm 0.2$ and $F_{\gamma} = (9.5 \pm 3.0) \times 10^{-9}$ photons cm$^{-2}$ s$^{-1}$. The fit with a PLE gives $\Gamma = 1.4 \pm 0.6$, $E_c = 0.7 \pm 0.4$ GeV and $F_{\gamma} = (7.4 \pm 2.4) \times 10^{-9}$ photons cm$^{-2}$ s$^{-1}$.  These spectral parameters are consistent with those first reported in Tam et al. (2010). The likelihood ratio test gives $2\Delta \log($likelihood$) \approx 8.8$, indicating that the improvement by the PLE model has a significance of just below 3$\sigma$, noting that the PLE model has one more degree of freedom. Using 1.5 years of data, Tam et al. (2010) found that the two models fit the data equally well, due to the lower photon statistics.

After the state change, the best-fit parameters for the simple power law are $\Gamma = 2.3\pm 0.1$ and $F_{\gamma} = (8.9\pm 1.1) \times 10^{-8}$ photons cm$^{-2}$ s$^{-1}$. The best-fit parameters for the PLE model are $\Gamma =1.8 \pm 0.2$, $E_c = 2.3\pm0.9$ GeV and $F_{\gamma} = (7.4\pm 1.0) \times 10^{-8}$ photons cm$^{-2}$ s$^{-1}$. The likelihood ratio test gives $2\Delta \log($likelihood$) \approx 8.8$, indicating that the improvement by the PLE model has a significance of slightly below 3$\sigma$. There is approximately a factor of 10 increase in the photon flux after the state change, confirming an earlier report (Stappers et al. 2013).

Spectral points were obtained by fitting the normalization factors of point sources within 6\degr\ from the pulsar and the diffuse backgrounds in individual energy bins. Other parameters were fixed to the best-fit values obtained in the full energy band analysis. The 95\% c.l. upper limits were calculated in energy bins where J1023 had a significance below 3$\sigma$. 

\subsubsection{Temporal analysis}
\begin{figure*}
  \centering
  \includegraphics[width=0.9\textwidth]{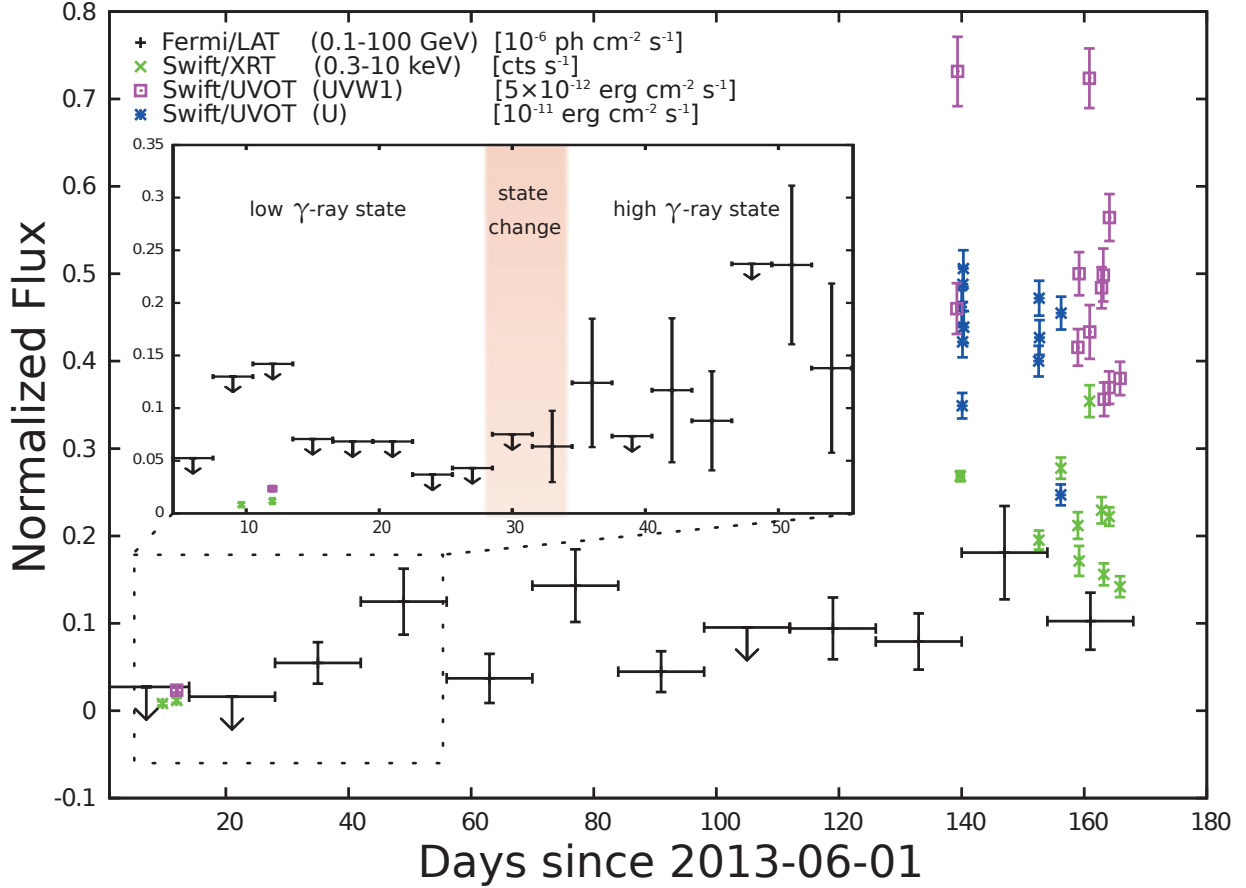}\\
  \caption{
The main panel shows the multi-wavelength (i.e., from UV to $\gamma$-ray) lightcurves of PSR J1023+0038 from June 1 to November 13 with different flux scales for each energy band (see upper left corner for details) while the inset box indicates the detailed evolution of the $\gamma$-ray emissions between June 6 to 24 July. UV/X-ray: Each datum represents an individual observation taken by Swift. $\gamma$-ray: Each datum in the main panel (inset) corresponds to two weeks (3 days), and 95 \% c.l. upper limits are given for the time intervals during which test-statistics (TS) values were below 9 (4).
}
\label{mwl_lc}
\end{figure*}

To show the change in $\gamma$-ray flux just before and after the state change, we constructed  $\gamma$-ray light curves since June 2013, as shown in Fig.~\ref{mwl_lc}, with the one shown in the inset focusing on the time interval of the $\gamma$-ray state change seen in late June/early July.  The same source model used in the aforementioned spectral analysis was used in deriving each flux value in the light curves, except that the power law model was assumed for J1023 due to the shorter time scales. The bin size is 14 days and 3 days for the main panel and inset, respectively. Upper limits were derived for those time intervals in which the TS values were below 9 (4) for the main panel (inset), assuming a power-law index of $\Gamma = 2.3$. It is clear that J1023 cannot be detected in two week's time before July 2013; we designate such time interval as {\rm low $\gamma$-ray state}. After the state change (a time interval we call {\rm high $\gamma$-ray state}), the $\gamma$-ray flux has remained at a much higher level for most of the time.

We estimated the dates of the state-change, i.e., when the binary system changed from the low $\gamma$-ray state to the high $\gamma$-ray state, as follows. Likelihood analysis were performed for the time intervals spanning one week (7 days). The first time interval that we used starts at June 20 and ends June 26, then we shifted the analyzed time interval for one day, i.e., the next time interval is hence June 21--27, the third time interval is June 22--28, and so on. The last time interval is July 8--14.

In the beginning, TS values are below 10. The first bin having TS$>$10 occurred at June~29--July~5. Note that never before (or, never since March 2013) could J1023 be detected at this significance with only one week of data, so during this particular week (June~29--July~5) J1023 has undergone a state change from the low state to high state in $\gamma$-rays. We indicate this time interval in Fig~\ref{mwl_lc}.

To look for orbital modulation, we mapped orbital phase to geocentric arrival time and divided the data into two halves in orbital phase, each centered at one of the conjunctions. We performed the same analysis as above in each phase interval and did not detect any significant orbital modulation.

\subsection{XRT Observations}
Since the disappearance of the radio pulsation of J1023, the pulsar has been monitored by \textit{Swift}/XRT for the X-ray evolution. The X-ray properties observed in 2013 June 10 and 12 (Kong 2013) are roughly consistent with previous {\it XMM-Newton} observations taken in 2004 and 2008 
(Archibald et al. 2010; Tam et al. 2010), indicating the system was still 
in quiescence. Since October 18, an enhanced X-ray emission was detected from 
a 10~ks XRT observation (Kong 2013; Patruno et al. 2014) reflecting a rapid X-ray 
activity in the binary system. Since then, a daily monitoring campaign was 
triggered to monitor J1023 almost daily in 
November (Pappito et al. 2013) and the X-ray source has 
stayed at a high flux level with high variabilities on timescales 
of a few tens of seconds. In this study, we mainly focus on the observations 
taken after the X-ray brightening (i.e., October 18) to investigate the 
origin of the enhanced X-ray radiations. All the \textit{Swift} observations
 used in this analysis are listed in the Table \ref{tab:swift}. 

We downloaded the observations from the \textit{Swift} Quick Look Archive and extracted spectra as well as lightcurves by the \texttt{HEAsoft} (version 6.14) \textit{Swift}-specific tasks \texttt{xrtgrblc} and \texttt{xrtgrblcspec} in which count-rate dependent regions are applied. 
As no significant spectral variability are seen among the individual observations during the high state, we combined all the selected events to construct an averaged X-ray spectrum (0.3--10~keV) with a total exposure time of 22.6~ks, which can be well described by an absorbed power-law model with best-fit parameters of \nh\ $=5.2^{+1.6}_{-1.5}\times10^{20}$\cm\ and $\Gamma=1.76^{+0.07}_{-0.06}$ ($\chi_\nu^2=0.95$ with dof $=170$), inferring an observed flux of $F_\mathrm{0.3-10}=(9.8\pm0.4)\times10^{-12}$\flux\ and an absorption corrected luminosity of $L_\mathrm{0.3-10}=(2.20\pm0.07)\times10^{33}\,(d/1.3\,\mathrm{kpc})^2$\lum. 
Comparing with previous {\it XMM-Newton} results (Tam et al. 2010), 
the X-ray luminosity is increased by at least a factor of 20 with a much 
softer photon index. We also tried to improve the fit by adding a thermal 
model such as \texttt{bbody} and \texttt{mekal} (\texttt{XSPEC} built-in 
models). Although it slightly improves the chi-square statistic, the F-test
 indicates the null-hypothesis probabilities for the \texttt{bbody} 
and \texttt{mekal} cases to be 14\% and 54\%, respectively suggesting that 
the improvement is not significant. In addition, the featureless X-ray
 spectrum does not favor any plasma emission contributions. 

As the average observed count rate is about 0.21 cts~s$^{-1}$, a 50-second bin time is used for the lightcurve binning to achieve a mean signal-to-noise ratio $>3$ per bin approximately. Short term variabilities on timescales of a few tens of seconds are clearly seen in the curve with variations up to a factor of ten between two consecutive bins. Interestingly enough, at least 21 of the data bins (with fractional exposure equals one) have zero count rate, which are unlikely random events due to an extremely low Poisson probability of getting a null count in a 50-second exposure (i.e., 0.0027\%). Some of the zero count bins are close enough with other low count rate bins ($\leqslant$ 0.07 cts~s$^{-1}$) to form several low-flux intervals in ranges of 200 to 550 seconds. 
Meanwhile, the peak flux reaches $1.64\pm0.23$ cts~s$^{-1}$ which arises from  $0.10\pm0.06$ cts~s$^{-1}$ in 100 seconds, stays for about 50 seconds, and then drops to $0.06\pm0.05$ cts~s$^{-1}$ in 100 seconds. 
However, no significant hardness variability is found despite the high flux variability. 
The \texttt{HEAsoft} tasks \texttt{powspec} and \texttt{efsearch} were employed 
to search for the signals from the known orbital and pulsar 
periods (Archibald et al. 2009; Wang et al. 2009). Although no reliable
 signal regarding to the orbit nor the pulsation is found, an unknown
 periodic signal of $P\approx245$s is detected from UT 6:36:42 to 7:04:35 on 
October 19 (about 7 cycles), which is not detected in other
 observations.

\subsection{UVOT Observations}
During the X-ray high state, \textit{Swift}/UVOT observed the source simultaneously with the XRT using the U (3465$\mathrm{\AA}$) and/or UVW1 (2600$\mathrm{\AA}$) band filters with IMAGE or EVENT mode (Table \ref{tab:swift}). The optical/UV counterpart is clearly detected in all UVOT images. Using the \texttt{HEAsoft} tasks \texttt{uvotmaghist} and \texttt{uvotevtlc}, we found that the UV magnitudes are much brighter than those quiescent measures taken in June by $\sim3.7$ mag. The UV emission varies among the observations between 16.76--15.98 mag (U) and 16.90--16.12 mag (UVW1) with uncertainties less than 
0.07 mag (extinction uncorrected). No trend can be found when folding the 
obtained lightcurve by the orbital period $P_\mathrm{obs}=0.198094(2)$ 
days (Thorstensen \& Armstrong 2005) suggesting that the variability is not 
orbit related. EVENT mode observations were started from November 6, which 
allow us to investigate short term variabilities with timescales down to a 
few tens of seconds. We chose a binning factor of 50 seconds and found that 
the binned lightcurves are full of variabilities with timescales between
 $\sim100$ and $\sim500$ seconds. The corresponding XRT data were examined
 but no clear correlation between the UV lights and the X-rays was found. 

\begin{table}[h]
\centering
\caption{\textit{Swift} Observations of PSR J1023+0038}
\scriptsize
\begin{tabular}{cccccc}
\hline
Obs Id & Date\footnote{All observations were taken in 2013. } & Exp & UVOT filter & Mode\footnote{UVOT observing mode. I: IMAGE mode; E: EVENT mode. } & X-ray state \\ 
& (m/d)  & (s) & (U/UVW1) & (I/E) & (\textbf{H}igh/\textbf{L}ow) \\ 
\hline
80035001 & Jun 10 & 1918 & UVW2 & I & L\\
80035002 & Jun 12 & 1863 & UVW1 & I & L\\
80035003 & Oct 18/19 & 9907 & U \& UVW1 & I & H\\
33012001 & Oct 31 & 1910 & U & I & H\\
33012002 & Nov 04 & 2083 & U & I & H\\
33012003 & Nov 06 & 1011 & UVW1 & E & H\\
33012004 & Nov 07 & 939 & UVW1 & E & H\\
33012005 & Nov 08 & 1264 & UVW1 & E & H\\
33012007 & Nov 10 & 1116 & UVW1 & E & H\\
33012008 & Nov 11 & 1134 & UVW1 & E & H\\
33012009 & Nov 12 & 2295 & UVW1 & E & H\\
33012010 & Nov 13 & 1121 & UVW1 & E & H\\
\hline
\\
\end{tabular}
\label{tab:swift}
\end{table}


\subsection{Keck/DEIMOS Optical Spectrum}
We obtained a 500-second spectroscopic observation from the DEIMOS multi-object spectrograph with a 0.8 arcsec slit on the Keck II telescope at UT 15:29:15 on November 3. 
The spectrum covers from 4750$\mathrm{\AA}$ to 9600$\mathrm{\AA}$ 
at $2.8\mathrm{\AA}$ resolution. The DEEP2 DEIMOS Data Pipeline was used for
 the CCD processing and flux was calibrated using HD19445 and the last 
photometric measure of Halpern et al. (2013) (i.e., $V=16.44$ mag). 
Due to an unstable weather condition during the observation as well as the large spectral deviation between the calibrator (Ap) and the target (G-type+low temperature accretion disk), we note that the flux calibration may not be very accurate.
Obvious emission lines of H$\alpha$, He \textsc{i} $\lambda$5876, and He \textsc{i} $\lambda$6678 are detected in the spectrum and a double peak feature is marginally seen in the H$\alpha$ line profile with a peak separation of $\sim19\mathrm{\AA}$, equivalent to 
a radial velocity difference of $\sim860\,\mathrm{km\,s^{-1}}$ that is 
consistent with the earlier spectroscopic analysis from Halpern et al. (2013). 
Similar to the SDSS spectrum taken during the last accretion state in
 2000--2001 (Wang et al. 2009), the continuum can be well fitted by
 a multi-temperature disk model plus an irradiated G5V companion 
of 17.35 mag given that the observation was taken during the plateau 
of the companion's modulated
 light curve (Thorstensen \& Armstrong 2005; Wang et al. 2009). 
Detailed accretion disk modelling will be discussed in the theory section 
(\S~\ref{theory}). 

\subsection{Lulin One-Meter Telescope}
On November 16, we observed J1023 with the Lulin One-Meter Telescope (LOT) at Lulin Observatory in Taiwan, using SDSS $g$, $r$, and $i$ filters. Two $g$-band (one minute and two minutes), six $r$-band (two minutes), plus one $i$-band (two minutes) exposures were obtained under a good weather condition. Standard data reduction and aperture photometry were performed by the \texttt{IRAF} packages to compute the mean magnitudes ($g=17.16\pm0.02$ mag, $r=16.63\pm0.01$ mag, and $i=16.42\pm0.03$ mag) for which the flux calibration was done by using three SDSS DR8 bright sources in the field (i.e., 
SDSS J102354.69+003516.2, J102357.97+004027.7, and J102338.46+003624.4) and the errors were calculated by assuming a Poisson distribution with standard propagation of errors.
The photometry is roughly consistent but less bluer than the DEIMOS spectrum. 
For the $r$-band images, we took the first image at 19:30 UT (near the time of inferior conjunction; see Thorstensen \& Armstrong 2005 for details) and consecutively made the rest of the observations (every 140 seconds) 36 minutes later. 
A large magnitude change of $\Delta m=0.15\pm0.03$ is seen from the first two images and small variabilities up to $\Delta m=0.09\pm0.02$ in 140 seconds are shown in the consecutive data set. 
For the $g$-band magnitudes, we transformed them from SDSS system to Johnson 
filter system (Jester et al. 2005) and compared them with the ones reported 
in Halpern et al. (2013).  The transformed $V$ magnitudes ($16.95\pm0.02$ and $16.88\pm0.02$) are 
very close to Halpern's values at similar orbital phase.

\subsection{Radio observations}
The radio emission from the position of J1023 was first reported 
by the FIRST VLA 1.4~GHz survey (Becker, White, Helfand 1995). In re-examining 
the FIRST data, Bond et al. (2002) realized that the detection reported by 
Becker et al. (1995) was based on a single $\sim6.6$~mJy flare from a 1.4~GHz 
observation on 10 August 1998. On the other hand, the observations on 
3 August 1998 and 8 August 1998 resulted in non-detections and the corresponding 
limiting flux densities are $<1.8$~mJy and $<3.4$~mJy, respectively 
(Bond et al. 2002). These observations clearly demonstrate the variability 
of the radio emission from this binary system, which can vary by a factor of 
$\sim4$ over a time interval as short as a week. However, without the optical 
spectroscopic results from this epoch, it remains uncertain whether the system 
was in a rotation-powered or an accretion-powered state. But transient radio 
emission from a neutron star in an accreting binary is possible 
(Gaensler, Stappers, \& Getts 1999). During the rotation-powered phase, 
the radio emission from J1023 becomes more intense. 
Archibald et al. (2009) reported a mean flux density of $\sim14$~mJy at 1.6~GHz. 
Recently, a long-term radio monitoring campaign of J1023
 from mid-2008 to mid-2012 reports the variable radio properties such as variable eclipses, short-term
disappearance of signal and excess dispersion measure at random orbital phases 
(Archibald et al. 2013). 

Based on a radio observation at 1.4~GHz performed on 23 June 2013, 
Stappers et al. (2013) reported the non-detection of pulsed radio emission 
from J1023 with a limiting flux density of $<0.06$~mJy. However, it should 
be cautious in interpeting the result as it does not necessarily provide 
evidence for the quench of coherent radio emission mechanism. For an alternative scenario, 
assuming the pulsar remains to be active, the matter transferred from the companion toward 
the neutron star can be ejected by the pulsar wind (Ruderman, Shaham \& Tavani 1989). 
Also, the matter evaporated from the disk by the pulsar can further complicate the 
circumstellar environment. Therefore, even with the presence of active 
pulsar mechanism, the non-detection of pulsed radio emission can possibly due to 
the enhanced scattering/dispersion in the environment which results in a smearing of 
the pulsed signal. Here, we build a theoretical model to explain the multi-wavelength observations of J1023 in its current accretion active state.

\section{Theoretical model}
\label{theory}

\begin{figure*}
\begin{center}
\rotatebox{-90}{\includegraphics[height=15cm,width=6cm]{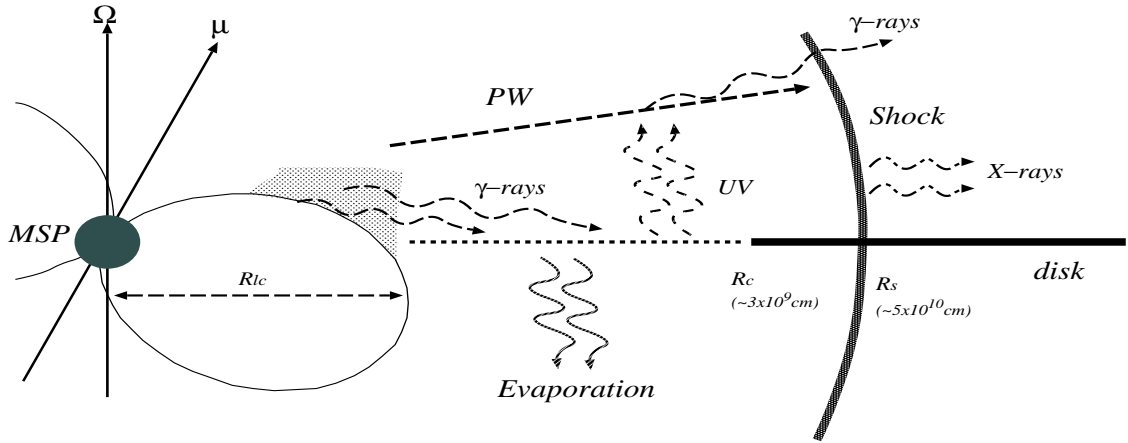}}
\caption{
The schematic view of multi-wavelength emissions from 
J1023 system after 2013 late-June. 
The accretion disk extends beyond the light cylinder radius ($R_{lc}$). $R_s$ is the distance to the intra-binary shock from the pulsar.
 Below the critical distance ($R_{c}$), the gamma-rays evaporate 
the disk matter.  UV/Optical photons are mainly produced by 
the disk emissions at $R\sim 10^{9-10}$cm. 
 The interaction between the pulsar wind and the stellar wind
 creates a shock and produces the non-thermal X-ray emissions.  
The inverse-Compton process of the cold-relativistic pulsar 
wind off UV/Optical photons from the disk produces the gamma-rays. 
}
\label{system}
\end{center}
\end{figure*}

In this section, we discuss a possible scenario for the emission processes 
in multi-wavelength bands after 2013 late-June. Figure~\ref{system} shows a 
schematic view of the system discussed in this section.  
With the observed flux of the UV emissions
 ($L_{UV}\sim 10^{33}{\rm erg~s^{-1}}$), the standard
gas pressure supported disk model (Frank et al. 2002)
implies a mass loss rate of 
 $\dot{M}_{16}=\dot{M}/10^{16} {\rm g~s^{-1}}\sim 1$ and the inner edge
of the accretion disk is $R_{in}\sim 10^{9-10}$cm from the pulsar.
This result suggests that the accretion disk does not substantially extend
 down to the pulsar  or the emissions from the disk below $10^{9-10}$cm is
 inefficient. Indeed, from our optical spectroscopy, the peak velocity of H$\alpha$ infers an outer disk size of $\sim 10^{10}$ cm. Takata et al. (2010) proposed that heating associated with     
   deposition of the gamma-ray radiation from pulsar magnetosphere onto the 
accretion disk evaporates the accretion disk of a LMXB in
 quiescent state.  The gamma-rays are absorbed via  the
 so-called pair-creation process in a Coulomb 
field by nuclei  (e.g. Lang 1999), if the disk column density 
exceeds  $\Sigma_c\sim 10^2 {\rm g~cm^{-2}}$. The energy conversion 
from the gamma-rays to the disk matter promotes the evaporation of the disk. 
 Applying  the standard  disk model, the critical axial distance 
from the pulsar, below which the gamma-rays can evaporate the disk, 
 is $R_c\sim 3\times 10^9 \alpha_{0.1}^{-16/15}{\dot M}_{16}^{14/15}\Sigma_{c,2}^{-4/3}$cm, where $\alpha_{0.1}$ is the viscosity parameter in units
of 0.1, ${\dot M}_{16}$ is the accretion rate in units
 of $10^{16}{\rm g~s^{-1}}$ and $\Sigma_{c,2}$ is the critical column
density in units of $10^2{\rm g~cm^{-2}}$.

We expect that below the critical radius,  the accretion disk will  move 
toward the pulsar with a lower accretion rate.
The mass loss from the disk due to the gamma-ray deposition reduces
 the rate of the viscous dissipation,
and the distance variation of  surface temperature of the disk  is
 more moderate than the one ($T_d\propto R^{-3/4}$) predicted by
the standard disk model, in which the accretion rate is constant
with the axial distance. Because it is not obvious the temperature
distribution of the accretion disk evaporation due to the irradiation
of the gamma-rays, we assume a constant temperature at  $T_d(R<R_c)=T_d(R_c)$
 as the first order of approximation, where $T_d(R_c)$ is 
the temperature at the critical distance and
is estimated by the standard disk model. Beyond the critical distance
$R=R_c$, we apply the standard disk model, and we choose the temperature
$T_d(R_c=10^{10}{\rm cm})\sim 8500$K, which corresponds to an accretion rate
of $\dot{M}_{16}\sim 1$, to explain the optical/UV data.

The existence of the GeV gamma-ray emissions even after the disappearance of the pulsed
radio emission suggests that
the accretion matter does not enter into the pulsar magnetosphere, and
the rotation powered activities are still turned on.  Hence, we assume
 that the  disk extends beyond the light cylinder
radius $R_{lc}=cP/2\pi\sim 8\times 10^6$cm.
The flux at the frequency $\nu$ from the disk is
\begin{equation}
F_{\nu}=\frac{4\pi h\nu^3 \cos i}{c^2D^2}\int_{R_{lc}}^{R_{out}}\frac{RdR}
{{\rm exp}^{h\nu/k_BT(R)}-1},
\end{equation}
where $D\sim 1.3$kpc is the distance to the system,
$h$ is the Planck constant, $c$ is the speed of light, $k_B$ is
the Boltzmann constant and $i\sim 40^{\circ}-50^{\circ}$ (Wang et al. 2009)
is the  Earth viewing angle measured from the direction perpendicular 
to the orbital plane. With the present assumption of the  
temperature distribution, the luminosity of the disk emissions
 below $R=R_c$ is comparable to or smaller than that above  $R=R_c$.

Tam et al. (2010) expected that the GeV emissions during the
pulsed radio state are originated from the outer
gap in the pulsar magnetosphere, because
 the inverse-Compton processes of the shocked pulsar
wind and the cold-relativistic pulsar wind could not explain the observed
flux level.     
The increase in the flux by a factor of 10 after the the disappearance of the pulsed radio emissions implies that an additional component emerges and
it dominates the magnetospheric component.
We propose that the inverse-Compton scattering
process of the cold-relativistic pulsar wind off the soft photons
from the disk produces the additional gamma-rays.
With the observed luminosity of the  UV emissions from the disk,
 the depth of the inverse-Compton
scattering process  is estimated to be of order of
$\tau_{ic}\sim n_s\sigma_Tr \sim  0.4(L_{UV}/10^{33}
 {\rm erg~s^{-1}})(E_s/3{\rm eV})^{-1}(r/10^9{\rm cm})^{-1}$,
 where $n_s=L_{UV}/4\pi r^2cE_s$ is the number density of the soft 
photons, $L_{UV}$ is the luminosity of optical/UV emissions from the disk, 
$E_s$ is the soft photon 
energy and $\sigma_T$ is the Thomson cross section. Hence, if the disk extends 
below $R\sim 10^9$ cm, the inverse-Compton of the cold-relativistic 
pulsar wind can produce the observable gamma-rays. 
 We note that the up-scattered photons  will be observed as a steady component
 with the orbital phase, because the accretion disk surrounds the pulsar. 

It is believed that the cold-relativistic pulsar wind forms near the
light cylinder and carries almost all of the spin down energy into the
inter-stellar space. We assume that the pulsar wind mainly consists of the
electrons, positrons and magnetic field, and that the energy distribution
of the particles forms a mono-energetic function.
The Lorentz factor ($\Gamma_{PW}$)
of the bulk motion of the pulsar wind is related to the magnetization
parameter ($\sigma$), which is the ratio of the Poynting flux to the kinetic
energy flux. Near the light cylinder, the magnetic energy dominates the
particle energy, that is $\sigma\gg 1$. Beyond the light cylinder, the
magnetic field may be dissipated so that the magnetic energy is converted
 into the particle energy of the flow (Croniti 1990; Lyubarsky \& Kirk 2001).
The Lorentz factor at the  distance $R$ is estimated  as
\begin{equation}
 \Gamma_{PW}(R)=\left(\frac{1+\sigma_L}{1+\sigma(R)}\right)\Gamma_{PW,L},
\label{gampw}
\end{equation}
where $\sigma_L$ and
$\Gamma_{PW,L}$ are the magnetization parameter and the Lorentz factor
at the light cylinder, respectively. Because of the theoretical uncertainties
of the evolution of the magnetization parameter beyond the light cylinder,
we assume a power law form of
\begin{equation}
\sigma(R)=\sigma_L\left(\frac{R}{R_{lc}}\right)^{-\beta},
\label{sigma}
\end{equation}
  where the index $\beta$ is the fitting parameter. 
In fitting the orbital dependent spectrum of the gamma-ray binary 
PSR 1259-63/LS2883, Kong et al (2011,2012) applied the above spatial 
dependent magnetization parameter to explain 
the observed data.  We calculated the inverse-Compton scattering process 
with an isotropic photon field,
\begin{equation}
F(E_\gamma, R)=\int \frac{\sigma_{IC}(R)c}{E_s}\frac{dN_s}{dE_s}(r)dE_s,
\end{equation}
where $dN_s/dE_s$ is the soft photon field distribution from the disk,
and the cross section $\sigma_{IC}$ is described by
\[
\sigma_{IC}(R)=\frac{3\sigma_T}{4\Gamma_{PW}^2(R)}
\left[ 2q{\rm ln}q+(1+2q)(1-q)+\frac{(\Gamma_qq)^2(1-q)}{2(1+\Gamma_qq)}\right],
\]
where $\Gamma_q=4\Gamma_{PW} E_s/m_ec^2$, $q=E_0/\Gamma_q(1-E_0)$ with
$E_0=E_{\gamma}/\Gamma_{PW} m_ec^2$ and $1/(4\Gamma_{PW})^2<q<1$
(Blumenthal \& Gould 1970).  The isotropic soft-photon field
 is a good approximation because the solid angle of the disk measured from 
the pulsar wind is not small, and the inverse-Compton processes with 
various collision angles contribute to the emissions.  We also take into account  the anisotropic 
inverse-Compton scattering process of the pulsar wind off  the stellar
 photons, although its contribution to the emissions after 2013 late-June 
is negligible.

\begin{figure*}
\begin{center}
\includegraphics[width=17cm]{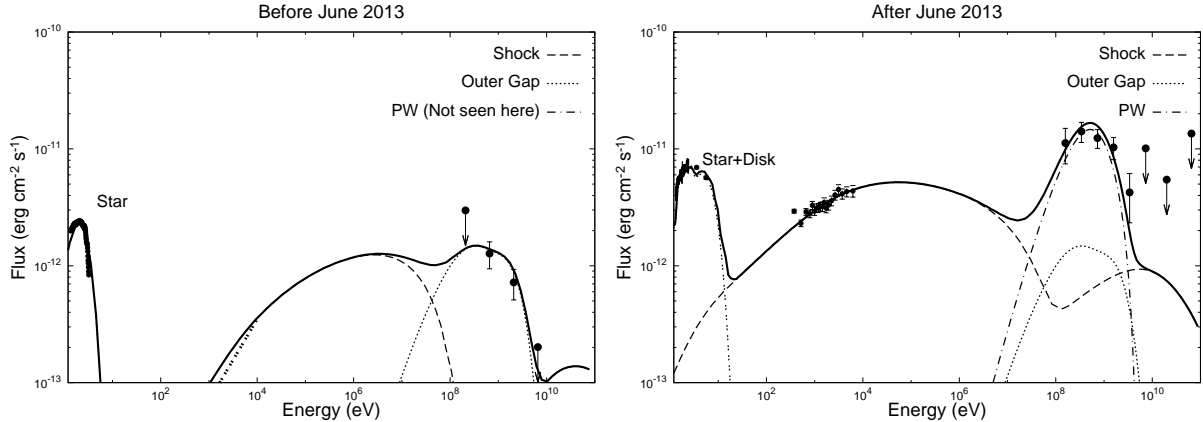}
\caption{Multi-wavelength spectra before (left) and after (right) 
2013 late-June. The dashed, dotted and dashed-dotted line
represent the calculated spectra of the emissions 
from the shock, outer gap (c.f.  Wang et al. 2010) 
and cold-relativistic pulsar wind, respectively.  
For the shock emissions, we assume  the distance  
$R_s\sim 8\times 10^{10}$~cm and the power law index $p=1.6$ 
 before 2013 late-June and $R_s\sim 5\times 10^{10}$cm and $p=2.2$ after 
2013 late-June, respectively.  
In the left panel, the stellar (G5V) spectrum (thick-curve line) 
and X-ray data (thick-double-dotted line) are taken from Wang et al. (2009)
 and Tam et al. (2010), respectively. The flux predicted by 
inverse-Compton process of the cold-relativistic pulsar wind 
before 2013 late-June is of order of $\sim 10^{-13} {\rm erg~cm^{-2}~s^{-1}}$
 and  its spectrum is not seen in the figure.}
\label{fit}
\end{center}
\end{figure*}

After 2013 late-June, the X-ray emission does not show the orbital variation and
increases by at least a factor of $\sim$20. Previous X-ray observations during the radio millisecond pulsar phase revealed orbital flux variation, implying the emissions
from the intra-binary shock (Tam et al. 2010; Bogdanov et al. 2011).
 Bogdanov et al. (2011) suggested that the emission region near
  the companion star and its orbital variation are caused by
the eclipse of the emission region by the companion star.
The disappearance  of  the orbital variation after 2013 late-June
 may suggest that the size of the  emission region is bigger
than that of before 2013 late-June.
We expect that the increase in the mass loss rate from the companion
star pushes the emission region  back toward the pulsar, and more fraction
of the pulsar wind is stopped by the shock, resulting an
 increase in  the X-ray emissions from the system.
The momentum ratio ($\eta$) of the stellar wind
and the pulsar wind is $\eta\sim (\dot {M}_wv_wc/L_{sd})\sim 1({\dot M}_w/10^{16}{\rm g~s^{-1}})
 (v_{w}/10^8{\rm cm~s^{-1}})(L_{sd}/5\cdot 10^{34}{\rm erg~s^{-1}})^{-1}$, indicating
the intra-binary shock stands at a distance
$R_s\sim a/2\sim 5\times 10^{10}$cm from
the pulsar if the mass loss rate of
the stellar wind after 2013 late-June increases to
$\dot {M}_w\sim 10^{16}{\rm g~s^{-1}}$.
With the momentum ratio $\eta$, the fraction ($\zeta$) 
of the pulsar wind stopped by 
the shock is estimated as $\zeta\sim (1-{\rm cos}\psi)/2$, where 
$\psi\sim 2.1(1-\eta^{2/5}/4)\eta^{1/3}$ (Eichler \& Usov 1993). 
Before 2013 late-June, because we do not know the mass-loss rate from the 
companion star, we assume that the X-ray emission regions is close to 
the L1-Lagrangian point and the size of the emission region is approximately 
described by the size of the Roche-lobe. The solid angle of the Roche-lobe around the companion star
measured from the pulsar is of order of $\delta\Omega\sim \pi (R_*/a)^2$ with
$R_*\sim 2\times 10^{10}$~cm being the radius of the Roche-lobe and  $a\sim 10^{11}$
the separation between two stars. 
Hence, the fraction ($\zeta$) of the pulsar wind stopped by the shock 
 will increase from  $\zeta\sim \delta\Omega/4\pi=1$\% before 2013 late-June 
to several ten percents after 2013 late-June. 

We expect  that a strong shock forms if the magnetization
 parameter is smaller than or comparable to unity $\sigma(R_{s})\le 1$ and
 the shocked pulsar wind particles form a power law distribution
described by $f(\gamma)=K\gamma^{-p}$. The maximum Lorentz
factor ($\gamma_{max}$) is determined from the condition for which
the particle acceleration time scale $t_{ac}\sim \gamma_{max} m_ec/eB$ is equal
to the synchrotron radiation loss timescale,
 $\tau_{sync}\sim 9m_e^3c^5/e^4B\gamma_{max}$, where
$B=3[L_{sd}\sigma(R_s)/R_s^2c(1+\sigma(R_s))]^{1/2}$ is the magnetic field
 just behind the shock. The minimum Lorentz factor and the normalization factor are calculated from the two conditions that $\int f(\gamma)d\gamma=\zeta 
L_{sd}/[4\pi\Gamma_{PW} (r_{s})r_{s}^2 m_ec^3(1+\sigma(r_{s}))]$ and
$\int \gamma f(\gamma)d\gamma=\zeta 
L_{sd}/[4\pi r_{s}^2 c(1+\sigma(r_{s}))]$,
 where $r_{s}$ is the radial distance to the shock  from the pulsar.
With a simple one-dimensional model, we solve
 the distribution of the pulsar wind particles with the cooling processes
via adiabatic expansion and radiation losses. We  calculate
the synchrotron radiation and the inverse-Compton radiation process from
the shocked pulsar wind.

Figure~\ref{fit} summarizes the results of the model fitting for
the observed emissions
before (left panel) and after (right panel) 2013 late-June.
 The dotted, dashed and dashed-dotted
lines are the calculated spectra for the emissions from the outer gap
(see Wang et al. 2010 for a detailed calculation), shocked pulsar wind
and the cold-relativistic pulsar wind, respectively. The double-dotted line
in the right panel shows the predicted spectrum of optical/UV emissions with
the current model of the accretion disk. The flux predicted by
inverse-Compton process of
the cold-relativistic pulsar wind before 2013 late-June is of order
of $\sim 10^{-13} {\rm erg~cm^{-2}~s^{-1}}$ and  its spectrum
is not shown in the figure.
As we can see in the figure that the observed GeV emissions 
after 2013 late-June
can be explained well by the inverse-Compton scattering process of
the cold-relativistic pulsar wind off the soft photons from the disk. We choose 
the parameters  of  the cold-relativistic pulsar wind described
by equations~(\ref{gampw}) and~(\ref{sigma}) as 
$\sigma_L\sim 100$, $\Gamma_{PW,L}\sim 200$ and the index $\beta\sim 0.7$, 
respectively. The observed flux level of the X-ray emissions
 after 2013 late-June is also reproduced  by the shock emissions with
a shock distance $R_{s}\sim 5\times 10^{10}$ cm and a power law 
index $p\sim 2.2$. Before 2013 late-June, a harder energy distribution ($p=1.6$)
  of the particles at the shock is required to explain the observed photon 
index ($\sim 1.4$) of the X-ray emissions. 

The matter evaporated from the disk  will smear out the pulsed radio emissions
from the neutron star via scattering and/or absorption, if the local
plasma frequency is greater  than the frequency of the radio wave.
 Assumed that the disk matter below $R_c\sim 3\times 10^9$cm are 
evaporated by the gamma-rays, one may estimate the local 
plasma frequency at the scattering region as 
$(\omega_p/2\pi)\sim 6{\rm GHz}(\dot{M}_d/10^{16}{\rm g~s^{-1}})^{1/2}
(v_{d,w}/10^8{\rm cm~s^{-1}})^{-1}(R_{c}/3\cdot 10^{9}{\rm cm})^{-1}$, where 
$\dot{M}_d$ is the evaporation of the  disk matter, and 
$v_{d,w}$ is the velocity of the evaporated matter. This suggests that 
the matter evaporated from the disk smears out the pulsed radio emissions
 up to a frequency of several GHz, which is consistent with observations that no radio pulsation has been detected below 5 GHz during the accretion active state. In this scenario, the rotation-powered MSP is still active. While detecting the pulsation can be difficult, radio imaging can still be used to check the aforementioned scenario as such technique is not hampered by the complicated environment. Also, as the local charge density can possibly be increased 
by the aforementioned processes, if the observed frequency is below the local plasma density, 
this can lead to a non-detection even with imaging. In view of this, a high frequency 
radio imaging observation is encouraged to confirm/refute this alternative scenario.

\section{Discussion}
\label{dicussion}
We performed a multi-wavelength study of the MSP binary J1023 using UV/optical, X-ray, and gamma-ray data during its accretion active state starting at late June 2013. Not only there are significant UV/optical and X-ray enhancements, the gamma-rays have increased by a factor of $\sim 10$.  We suggest that there is a newly formed accretion disk associated with UV emission and the enhanced gamma-rays are produced via inverse Compton scattering
between the UV emission from the disk and pulsar wind. At the same time, some fraction of the pulsar wind is stopped by the intra-binary shock resulting an increase in X-ray radiation. Because of the gamma-rays from the pulsar magnetosphere, the accretion disk has been evaporated, smearing out the pulsed radio emission even though the rotation powered MSP is still active. 

As Figure~\ref{mwl_lc} indicates, the gamma-ray emission has brightened by a factor of 10 on a timescale of several days. 
In the inverse-Compton scenario discussed in \S~\ref{theory},
 the brightening timescale seen by $Fermi$ 
corresponds to a migration timescale of the disk down 
to $r\sim 10^{8-9}$cm, where the inverse-Compton scattering process 
 is more efficient with a shorter mean free path than the process
 occurred at outer part of the disk
 $R\sim 10^{10-11}$ cm. If one assumes a standard disk model,
 one can see that  the migration timescale of the disk from $R\sim 10^{10}$cm 
 to $R\sim 10^{8-9}$cm is of order of several to ten days, that is, 
$\tau_{mi}\sim R/v_{R}\sim 9\alpha_{0.1}^{-4/5}
M_{16}^{-3/10}R_{10}^{3/4}$days, where $v_{R}$ is the radial velocity of 
the disk and $R_{10}=(R/10^{10}{\rm cm})$. 

Because  UV/optical  emissions are originated 
from the accretion disk, their variability timescale $\tau_{UV/opt}\sim 100$s
may correspond to the  Keplerian motion  at the emission 
region. The standard disk model implies that
 UV/optical emissions with the observed luminosity level are mainly
 from the axial distance $R\sim 10^{9-10}$cm, corresponding to 
an orbital period of $\sim 160(R/5\cdot 10^{9})^{3/2} {\rm s}$, which is 
 consistent with the observed timescale.

 In the X-ray energy bands, the origin of the timescale 
$\tau_X\sim 50-100$s of the observed variability
 is more speculative with the shock emission  model discussed 
in \S~\ref{theory}. But one may consider that the observed variability 
is caused by either  perturbation of shock front due to clumpy stellar wind or 
wind speed variation ($v_w\sim 10^8{\rm cm~s^{-1}}$), or sound propagation 
($c_s\sim 10^8{\rm cm~s^{-1}}$) in the shock front ($R_s\sim 5\times 10^{10}$cm).
 A more detailed consideration will be required to identify the origin of the
 X-ray variability.

JT and KSC are supported by a GRF grant of HK Government
 under HKU7009 11P. AKHK is supported by the National Science Council
 of the Republic of China (Taiwan) through grants NSC100-2628-M-007-002-MY3, NSC100-2923-M-007-001-MY3, and NSC101-2119-M-008-007-MY3. PHT is supported by the National 
Science Council of the Republic of China (Taiwan) through grant NSC101-2112-M-007-022-MY3. CYH is supported by the National Research Foundation of Korea through grant 2011-0023383.

\end{document}